# EARLY PERFORMANCE PREDICTION OF WEB SERVICES

Ch Ram Mohan Reddy[1], D Evangelin Geetha[2], KG Srinivasa[2],

T V Suresh Kumar[2], K Rajani Kanth[2]

[1]B M S College of Engineering, Bangalore - 19, India
[2]M S Ramaiah Institute of Technology, Bangalore - 54, India

crams19@yahoo.com, degeetha@msrit.edu, kgsrinivas78@yahoo.com,
tvsureshkumar@msrit.edu, rajanikanth@msrit.edu

## ABSTRACT

*Web Service is an interface which implements business logic. Performance is an important quality aspect of Web services because of their distributed nature. Predicting the performance of web services during early stages of software development is significant. In this paper we model web service using Unified Modeling Language, Use Case Diagram, Sequence Diagram, Deployment Diagram. We obtain the Performance metrics by simulating the web services model using a simulation tool Simulation of Multi-Tier Queuing Architecture. We have identified the bottle neck resources.*

## 1. Introduction

Web services are modular, self-contained "applications" or application logic. These services are developed per a set of open standards. There is even concurrence of this application-centric viewpoint from the World Wide Web Consortium (W3C). With the current Web Services Architecture specification, which states categorically that a Web service is indeed a software system? This stake in the ground from W3C goes a long way toward Non-Functional requirements such as reliability, maintainability, availability, security and performance, etc.

A Web service typically not meant to be a full-blown, feature-rich application in its own right. First, Web services need not to have a user interface. This alone runs counter to most people's concept of what constitutes an application. A Web service is better thought of as a "mini-application," possibly even an application "segment," or better still as an application enabler. However, to assess preliminary-early assessment of Non-functional requirements it may be useful to consider the prototypes/user interface for a typical web service such as eBay.

Figure 1 illustrates a future web application. This hypothetical web application is intended to provide a "wish you were there" vacation planning and vacation reservation function. To achieve this objective, this application is shown gainfully exploiting multiple web services, from disparate sources, to synthesize a highly compelling, very up-to-date full-function, and seamless user experience. The rich and topical functionality, available in the form of Web services from third parties, ensures that this application can be flexible, extensible, easily modifiable, and, above all, highly effective. These objectives make the assessment of non-functional requirement such as performance more complex. Towards assessing the performance for applications of this kind with the screen available on the internet is quite complex as we are not aware of the underlying software architecture. Especially non-functional requirements play the dominant role. Performance refers to the system responsiveness with respect to the time required to respond to specific events, or with the number of events processed in a given time interval. Performance is vital for software systems that perform customer-service functions that must provide a rapid response which is acceptable by the customers. Unfortunately, performance is the neglected aspect of the software-development methodologies. The increasing complexity of the software systems makes it essential that the performance aspects to be analysed from the early phases of the software life cycle [4], [19]. Examining the performance at the end of software development is a common practice in the industry, but it may lead to expensive and powerful hardware usage, time consumption in tuning measures or in the worst case, complexity redesigning the application.

A UML use case is a behavioural classifier. From a performance perspective, use cases are used to identify the workloads that are significant from a performance point of view that is the collections of requests made by the users of the system. . Usage of UML is well-known in SPE research community. Since UML is widely used as modelling language [12], [16], [23].

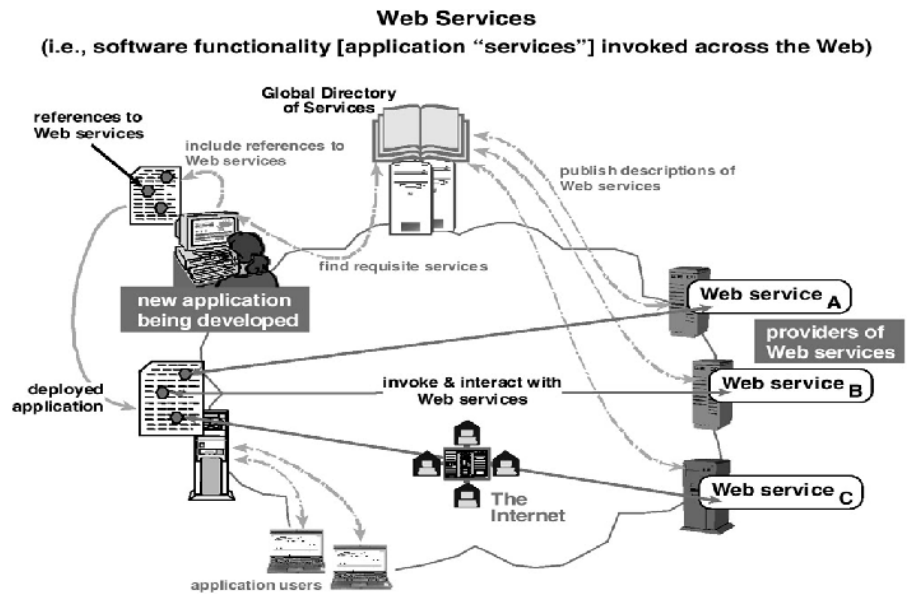

Figure 1. Planning and vacation reservation application of Web Services [1].

Web service is an interface implements the business logic as a group of operations which is accessible through various Internet Protocols [13]. It abstracts the information about the service and permits for utilizing the service from which it is implemented. The basic web service architecture [15] is based up on the communication between three roles. Service provider, service broker and service requester. These three perform to publish, find and bind operations. Figure 2 shows the web service described using standard XML notation called service description using WSDL [2]. It represents the necessary information to interact with service. SOAP [21] helps in organizing and interacting process using XML for information transmission between peers. UDDI [24] is the discovering technology used to find the exact web service according to request.

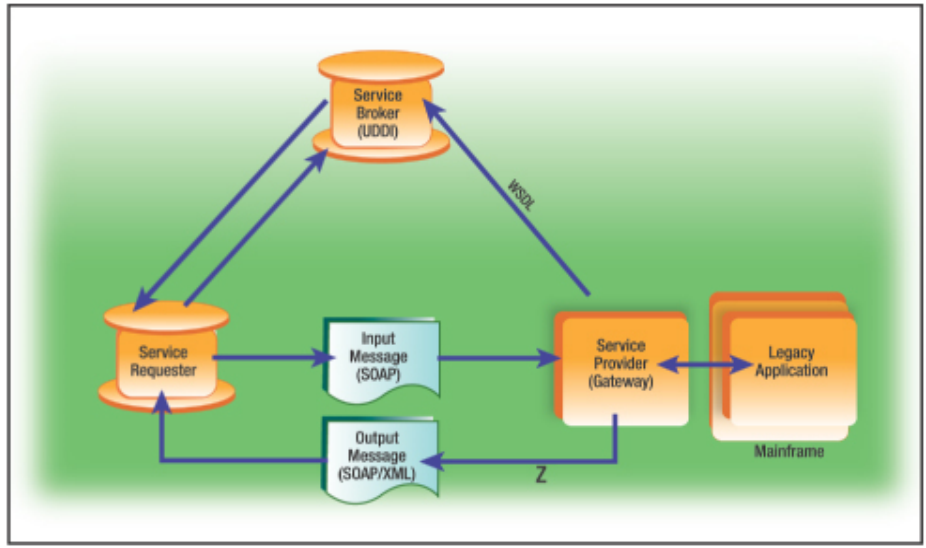

Figure 2. Web services architecture [28].

Performance is a one of the major problem in Web Services. Web service frameworks do not include the functionality required for web service execution performance measurement from an organization perspective. Here it is essential to address the performance is full in the context of Web Services. It is also important to discuss performance prediction techniques which are useful in reducing the cost for development. We use Software Performance Engineering (SPE) for Web Service.

SPE is a methodology to predict the performance of software systems early (analysis phase) in the life cycle [4]. SPE continues through the detailed design, coding and testing stages to predict and manage the performance of the evolving software and to monitor, report actual performance against specifications and predictions. SPE is important for software engineering and in particular for software quality. The Software Performance Engineering Process uses multiple performance assessment tools depending on the state of the software and the amount of performance data available.

SMTQA (Simulation of Multi-tiered Queuing Applications) is a process-oriented simulation tool, developed for the performance evaluation of software that follows multi-tier architecture [11]. It provides full visualization of model structure, parameters and output reporting. It addresses the following issues in a distributed environment.

- Simulate the multi-tier architecture with open workload and multi-classes

- Consider the software resource requirements of Use case Performance Engineering (UPE) approach.

- Simulate the behavior of the servers with replicas

- Balance the workload among the replicas using dynamic load balancing algorithm

- Obtain the performance metrics such as server utilization, average response time, average waiting time, average service time, probability of idle server and probability of dropping of requests required for capacity planning

- Generate the graphs for analysing the performance metrics.

## 2.     Related Work

Web services mainly used for the purpose of communicating with each other and clients. Web services can transfer the data to others behind firewall. Web service shares business logic, data and processes through interface over network. Now developers can add this service to their GUI to offer that service. Web services permits numerous applications from several resources to communicate. Web services are not tied to any operating system or any programming language.

Meeting Performance Requirements is an important issue towards acceptability of given software. The performance of the SOA is primarily studied as web service performance, since the web service is one of the main supporting technologies of SOA. A prototype service oriented application has been executed, and the actual performance is measured in [25]. Kohl Hoff's work concentrations on the performance evaluation and analysis of the SOAP protocol in a web service [8]. Spitznagel and Garlan have used queuing network for analysing simple client-server system in [5]. Garlan et al. demonstrated how formal approaches to software architecture can lead to boosts in software quality, including improved clarity of design, provision for analysis, and assurance that implementations conform to their intended architecture in [9]. Gamble's work focuses on detecting architectural discrepancies between web services by making a minimal web service architectural style [6].

Due to the intricacy of processors and micro architectures, simulations are used to calculate their performance [14]. In the context of processor simulation two approaches exist, i.e. trace-driven simulation and execution-driven simulation. Trace-driven simulation uses captured or unnaturally generated trace files as input and simulates their timing behavior on a modeled system. This approach is an old technique and widely used [14], [18]. Execution-driven simulation uses software programs as input and simulates their functional execution. Simple Scalar [27], [10] is an example for this approach. The execution-driven approach suffers from the drawback of a fix instruction set and the necessity to port operating systems and drivers to the simulation framework. In this paper, we have used model based simulation, since the prediction of performance is done during feasibility study of Software Development Life Cycle (SDLC).

Software performance engineering methods (SPE) [3] use interpreted Unified Modelling Language (UML) diagrams to model the system and soft-ware under study [17], [20], [26]. Since UML does not allow for the modeling of nonfunctional aspects many authors apply the UML Profile for  schedulability, Performance, and Time Specification (SPT) [22] to enhance the diagrams with the necessary semantics [17], [26]. The UML Profile for Modeling and Analysis of Real-Time and Embedded systems (MARTE) is the successor of the SPT profile, allows for a detailed modeling of performance aspects, and supports UML 2.

We have modelled the functionalities of web services in general using UML 2.0 use case diagram and sequence diagram.

## 3.     Methodology

To assess the performance of web services, we have exploited the methodology given in [use case point]. The web services proposed steps involved in prediction are as follows.

1. Develop the use case model for the general web services.

2. Generate the sequence diagrams, for representing the flow of events in each use case.

3. Consider the execution environment of the software components.

4. Simulate the model using SMTQA and obtain performance metrics.

5. Analyse the performance metrics to achieve the performance goal.

## 4. Application of Methodology

The Use Case model of Web Services and the corresponding sequence diagrams are given in figure 3 and figure 4 respectively.

Four steps involved in the process of engaging a Web service are:

a. The requester and provider must be known to each other to initiate communication between them. There are two cases.

    1  In a typical case, the requester who would initiate the process must be aware of service provider.

       There are two ways:

- The requester may get provider's address directly from the provider.

- The requester may use a discovery service.

    2  In other case, the provider may initiate communication between requester and provider. They would get to know each other wherein the provider agent somehow obtains address of the requester agent.

b. If the requester and provider agree on service description (a WSDL document), policy constraints would make successful communication between requester and provider. However, it does not assure that requester and provider communicate with each other. It simply suggests that both must have the same policies of the service description.

    There are different ways this could be achieved:

- Through direct communication between requester and provider, to know the service description and polices.

- Requester must accept the policies of service provider.

- Both requester and provider must follow an industry defined standard.

- Service description and polices defined and published by the requester and offered to provider.

c. The service description and polices are input to, or embodied in, both the requester and the provider as required.

d. SOAP messages can be communicated between requester agent and provider agent on behalf of their service provider and service requester.

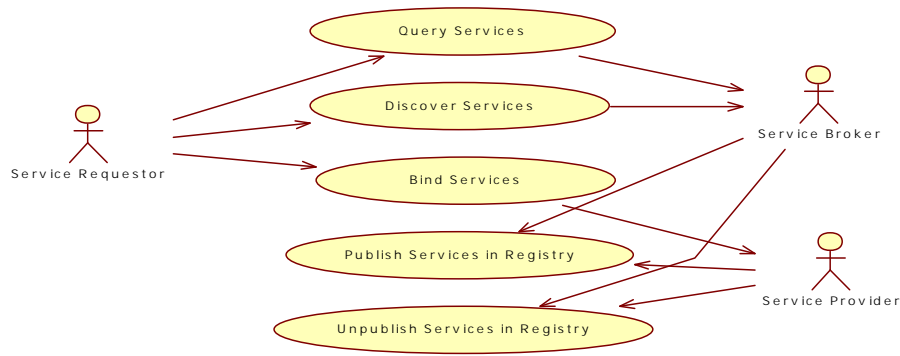

Figure 3. Use Case model for Web Services.

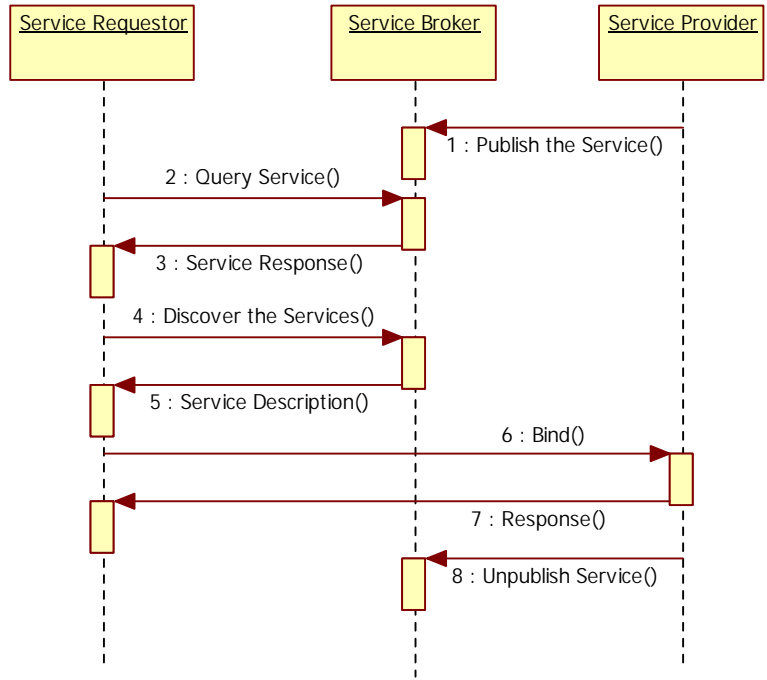

Figure 4. Sequence Diagram for Web Services.

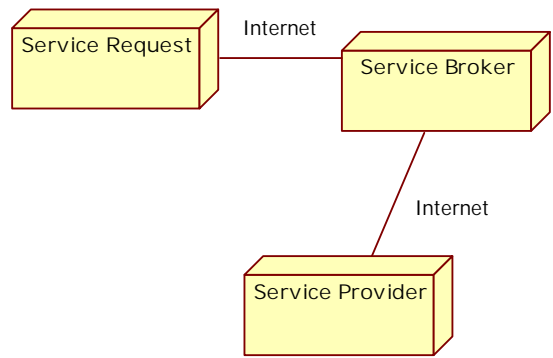

Figure 5. Deployment Diagram for Web Services.

# 5. Simulation Results

The behaviour of the hardware resources can be analysed by simulation. Since the Web Services Technology is layered in nature, the simulation model is designed as multi-tier architecture. The model is simulated using the tool SMTQA [11]. The model is simulated for 1000 requests. 3-tier architecture is considered for simulation. The layers considered are service requests, service broker and service provider. The inputs required for simulation are: software resource requirements, execution environment, software execution structure, and resource usage. The software resource requirements, i.e. software size, input size, web page size etc. are obtained using the methodology given in [7]. The sequence diagram given in Figure 4 is considered for software execution structure and the deployment diagram presented in Figure 5 is considered for execution environment. The performance metrics; resource utilization, average response time, average service time, average waiting time, probability of idle server and probability of dropping of requests are obtained. The graphs are generated for the performance metrics against arrival rate. For discussion, the graphs generated for average response time against arrival time obtained in resources of the tiers are presented in Figure 6 through Figure 10. From these figures, we could observe that in Internet 1 and Internet 2, the response time is high compared to other resources. The obtained performance metrics for web services is presented in Table 1. From this table, we could observe that Internet 1, Internet 2, Service Broker Disk are the bottleneck resources, due to the performance metrics, the Average Waiting Time and Probability of dropping of sessions.

Table 1 Performance Metrics obtained for Web Services using SMTQA

|  | Average Response Time | Average Service Time | Average Waiting Time | Probability of Idle Server | Probability of dropping of Sessions |
|---|---|---|---|---|---|
| SRS | 0.00012 | 0.00012 | 0.000 | 0.013 | 0.000 |
| Internet 1 | 0.207 | 0.057 | 0.150 | 0.003 | 0.761 |
| SBCPU | 0.00002 | 0.00002 | 0.000 | 0.044 | 0.000 |
| SBDsk | 0.016 | 0.016 | 0.000 | 0.040 | 0.000 |
| Internet 2 | 0.291 | 0.077 | 0.214 | 0.000 | 0.723 |
| SPCPU | 0.86 | 0.86 | 0.000 | 0.035 | 0.000 |
| SBDsk | 2.381 | 1.500 | 0.881 | 0.011 | 0.789 |

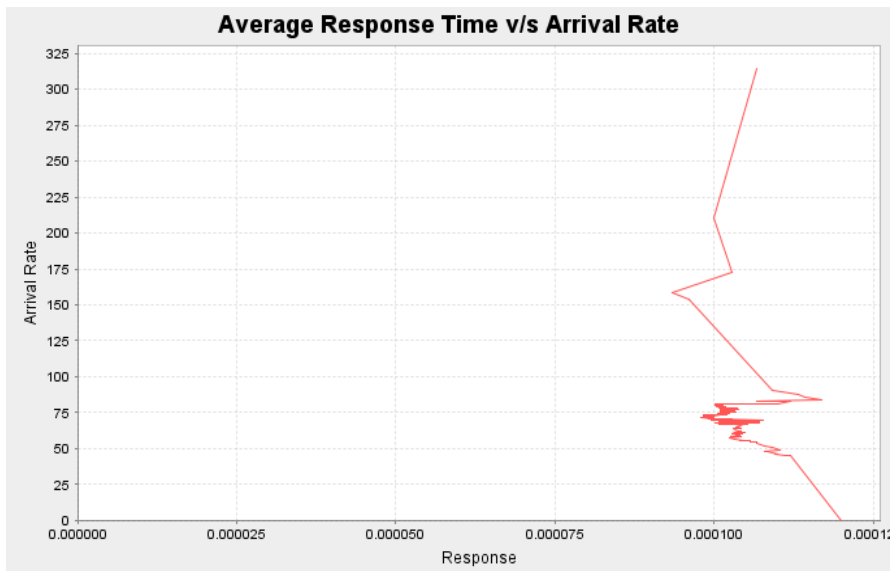

Figure 6. Service Request CPU.

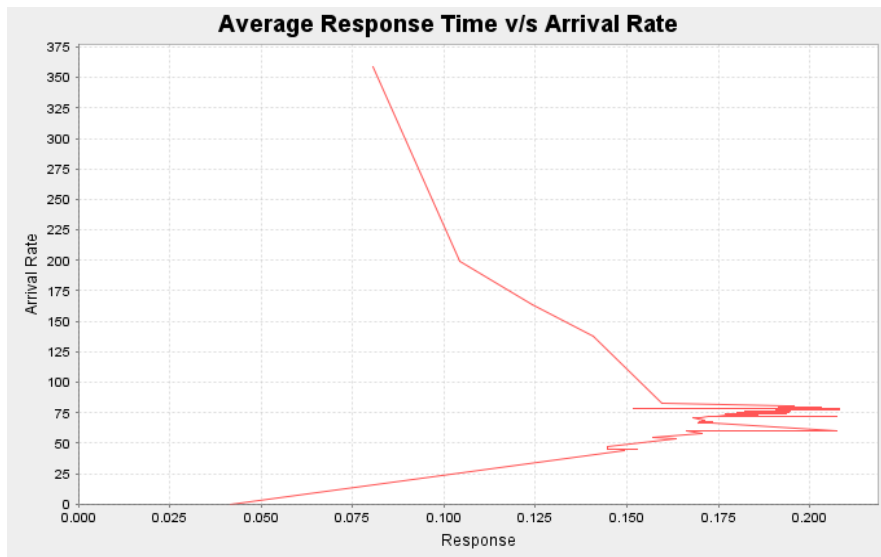

Figure 7. Internet 1.

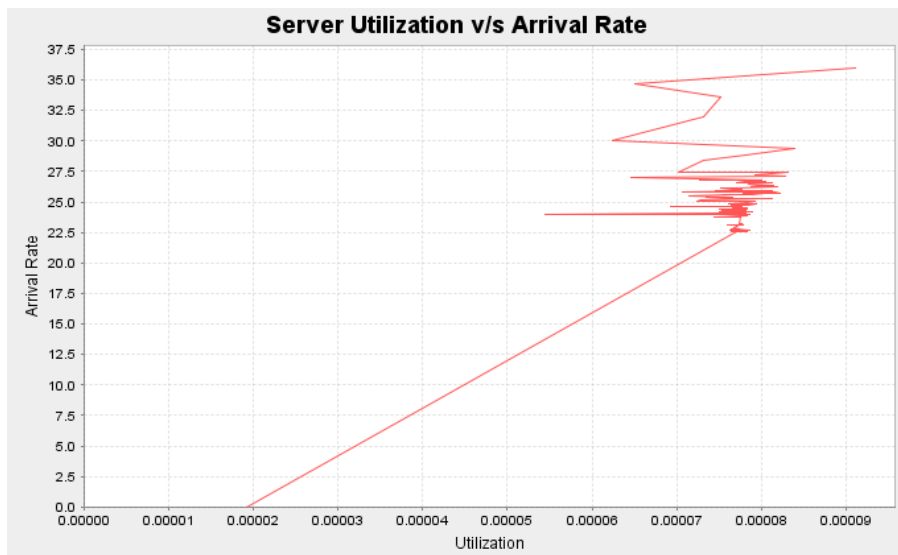

Figure 8. Service Broker CPU.

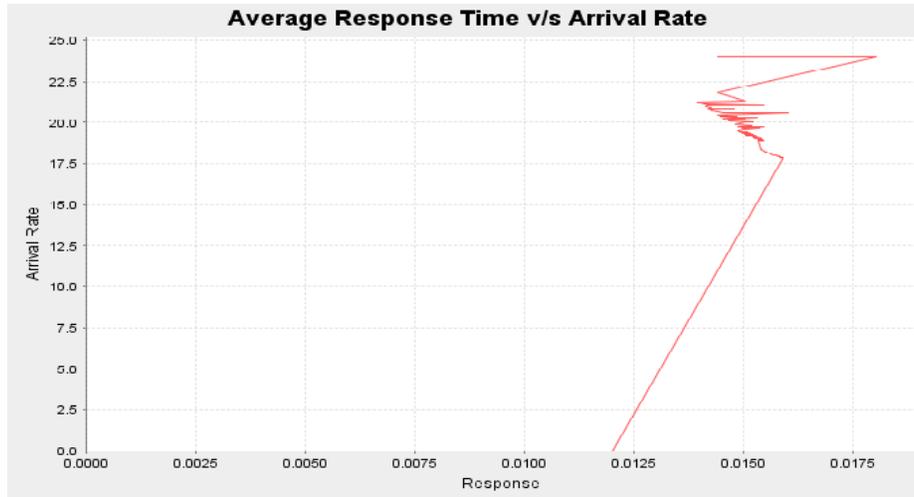

Figure 9. Service Broker Disk.

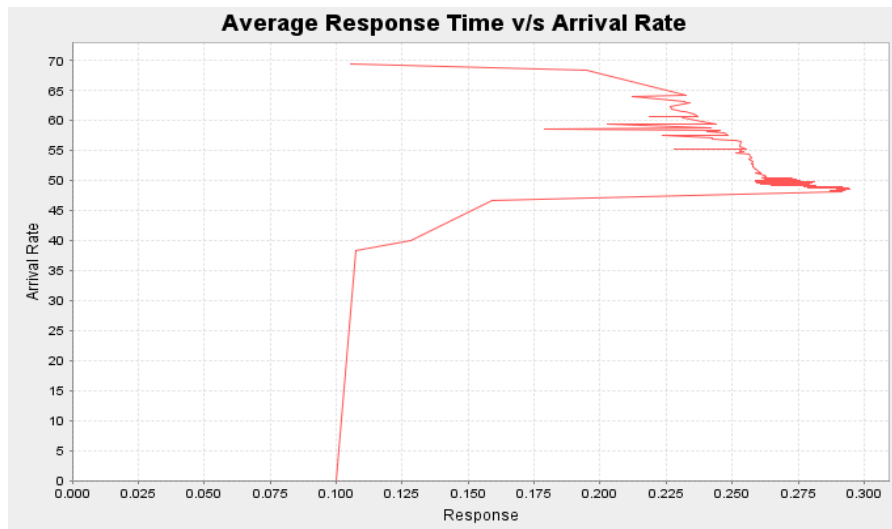

Figure 10. Internet 2.

From the graphs given in figure 6 to figure 10, we have observed the following:

In service request CPU, the response time is ranging between 0.00009 and 0.0002. The response time in internet is varying rigorously between 0.05 and 0.2. In service broker CPU, the response time is initially 0.00002 and gradually increasing up to 0.00008 and from that point onwards, it is varying between 0.00006 and 0.0009.

The response time in Internet 1 and Internet 2 is varying rigorously. In Internet 1, it is between 0.05 and 0.2 while in Internet 2, it is between 0.1 and 0.3. In service broker disk, it is varying between 0.0125 and 0.175.

# 6.    Conclusion and future work

Due to the dynamic behavior of Web Services, predicting response time during early phases of SDLC becomes complex. Hence, in this paper, we have modeled Web Services using UML models, Use Case diagram, Sequence diagram and Deployment diagram. The model is simulated using the tool SMTQA and the performance metrics are obtained. The response time obtained for the hardware resources are analysed, bottleneck resources are identified. In future, we plan to develop methodologies for software performance prediction for Web Services.